
\input jnl-0.3.tex


\font\titlefont=cmr10 scaled \magstep3
\def\bigtitle{\null\vskip 3pt plus 0.2fill \beginlinemode \doublespace
\raggedcenter \titlefont}

\gdef\journal#1, #2, #3, 1#4#5#6{
{\sl #1~}{\bf #2}, #3 (1#4#5#6)}

\def\ap{\journal Ann. Phys., }
\def\cqg{\journal Class. Quantum Grav., }

\def \e{e_\mu^a}
\def \w{\omega_\mu^{ab}}
\def \o{\omega}
\def \g{\gamma}

\preprintno{UCSBTH--91--40}

\bigtitle Ashtekar's Approach to Quantum Gravity$^*$
\body\footnote {}
{* To appear
in the proceedings of the Strings and Symmetries 1991 Conference held at
Stony Brook, May 20-25, 1991.}

\author Gary T. Horowitz

\affil \ucsb
\centerline{gary@voodoo.bitnet}

\abstract A review is given of  work by Abhay Ashtekar and
 his colleagues Carlo Rovelli, Lee Smolin, and others, which is
  directed at constructing a nonperturbative  quantum theory of  general
   relativity.

\endtitlepage

\baselineskip=16pt

\subhead{1. INTRODUCTION}

I have been asked to review the current status of an approach to quantum
gravity which is being developed by Abhay Ashtekar and his colleagues
Carlo Rovelli, Lee Smolin, and others\refto{rev}.
I should emphasize  that I have not actively worked on
this approach and as a result, my knowledge of it is somewhat incomplete.
However I
have followed the progress in this area and would like to describe for you
the main ideas involved, the current status, and open problems.

Ashtekar  and colleagues are trying to quantize standard four dimensional
general relativity
without supersymmetry, higher derivatives, extra dimensions, extended objects,
etc. The first question that probably comes to mind is why are they wasting
their time on a program that is doomed to failure? Isn't it well known that
general relativity cannot be quantized? Perhaps surprisingly,
the answer is no. It is, of course,
known that general relativity is not perturbatively renormalizable\refto{gs}.
But unlike the case for
most quantum field theories, this may not be as bad as it
sounds. General relativity is qualitatively different from other field
theories in that the dynamical field is the spacetime metric.
One might even argue that
standard field theory perturbation techniques
{\it should} break down  since they are
based on the assumption
that spacetime looks Minkowskian at arbitrarily short distances, which
is not very plausible in quantum gravity.  As I will describe, there
are indications that quantum general relativity provides a natural cut-off
at the Planck scale.

Ashtekar works in the framework of canonical quantization. Thus it is analogous
to the functional Schroedinger representation for ordinary field
theories\refto{fj}.
However, as we will see, the reparameterization invariance of general
relativity leads to certain simplifications. Canonical quantization of
general relativity has, of
course, been tried before. But previous investigations
have almost always used the spatial three metric and its conjugate momentum
as the basic canonical variables. This leads to constraints which
are difficult to solve (or even make sense of) in the quantum theory.
Ashtekar instead chooses canonical variables which are analogous to
those in ordinary gauge theories. The resulting constraints are simpler
and more progress can be made toward constructing the quantum theory.

To motivate Ashtekar's choice of canonical variables, we begin by considering
general relativity in three dimensions. This theory can be described
in terms of an SO(2,1) connection $\w$ and a triad of (dual) vectors
$\e$. (The spacetime metric is defined in terms of the triad by $g_{\mu\nu}=
 e^a_\mu e^b_\nu \eta_{ab}$.) The action is
$$    S = \int e^a \wedge R^{bc} \ \epsilon_{abc}  \eqno(1)$$
where $R = d\o + \o \wedge \o$
is the curvature
two form or field strength of $\o$. This action in fact describes a slight
extension of general relativity. When $\e$ consists of three linearly
independent vectors, one can show that (1) is equivalent to the usual Einstein
action $\int R \sqrt{-g}$.
But the action (1) and the resulting field equations remain well defined even
in the
limit that the triad becomes linearly dependent. Thus this theory includes
degenerate metrics. We will return to this point in Section 3.

Witten has shown\refto{wi3d} that if one chooses
the dynamical variables to be the spatial components of the connection $\o_i$
and its conjugate momentum $E^i$ (which is just the dual of the spatial
components of the triad), then
the canonical quantization of this theory can be
carried out exactly.
The theory has two constraints:
$$     D_iE^i=0 \eqno(2)$$
$$  R_{ij}=0 \eqno(3)$$
The first is the familiar Gauss' law. The second says that the spatial
connection $\o_i$ is flat. As a result of reparameterization invariance,
the Hamiltonian is proportional to the constraints.
Thus, to construct the quantum theory, one does not need to
solve the time dependent Schroedinger equation. It suffices to find states
which are annihilated by the quantum version of the constraints.
One can represent states in terms of functionals of the connection.
Imposing
(2) requires that $\psi(\omega)$ be gauge invariant and imposing (3) requires
that $\psi$ have support on just the flat connections. So physical
states are functionals of gauge inequivalent flat connections.

\subhead{2. TOWARD QUANTUM GENERAL RELATIVITY}

Since the above approach works so well in three dimensions,
it is natural to try it in
four. This leads directly to Ashtekar's variables. (Actually,
Ashtekar began his four dimensional work several
years before the three dimensional
case was considered\refto{aa}!) In
four dimensions, general relativity can be described in terms of an
SO(3,1) connection $\w$ and a tetrad of (dual) vectors $\e$. The action is
$$  S=\int e^a \wedge e^b \wedge R^{cd}\  \epsilon_{abcd} \eqno(4)  $$
Using the three dimensional case as a guide, one is tempted to consider the
spatial components of the connection and its conjugate momentum
as the basic dynamical variables.
Unfortunately, if one casts the theory into canonical form, one finds that
some of the constraints are now second class. One can explicitly solve the
second class constraints, but the remaining constraints become
nonpolynomial\refto{abj}.
Ashtekar's key insight\refto{aa,js1}
was to replace $\o_\mu$ with its self dual part
$A_\mu\equiv \o_\mu -i *\o_\mu$.
(One can show that the action (4) with $\o_\mu$ replaced by $A_\mu$ is still
equivalent to
general relativity.)
At first sight this appears to be a rather minor change. One is
essentially replacing a connection having 24
real components with one having 12 complex components.
However closer examination reveals that the consequences are
much deeper than that.

This is most easily seen
in the Euclidean context. The Euclidean Einstein equations can be
obtained from the action (4) with either an SO(4) connection or its
self dual part $A_\mu$ which is a {\it real} SU(2) connection. Thus one can
eliminate half of the components of the connection without losing any
information! (The reason is basically that the two actions differ by
a term proportional to $\int R_{[\mu\nu\rho\sigma]}$ which does not contribute.
Since one retains the full tetrad, one has the spacetime  metric and
in any solution, one can always
reconstruct the complete connection and its curvature.)
In addition
to the obvious economy of fields, there  are further advantages to
working with $A_\mu$.
For example one can show that Einstein's
field equation (with arbitrary cosmological constant) is equivalent to
the {\it self dual Yang-Mills equation for the connection $A_\mu$}\refto{hu1}.
(More precisely,
it is equivalent to the self dual Yang-Mills equation in a curved background
where $A_\mu$ is equal to the self dual part of the spin connection.) Using
this
correspondence, one can find gravitational analogs of SU(2) Yang-Mills
instantons: The one instanton solutions turn out to correspond to the
four sphere, with the size of the instanton related to the radius of the
sphere\refto{sa}.

Returning to the Lorentzian context, one finds further advantages of using
the self dual connection when one constructs the canonical formulation
of the theory. The dynamical variables are the spatial components of
the connection
$A_i$ and its conjugate momentum
$E^i$ which contains the information on the
tetrad. The constraints are all first class and take the form
 $$   D_i E^i = 0 \eqno(5)$$
 $$  {\rm Tr}\ F_{ij} E^i = 0 \eqno(6)$$
 $$  {\rm Tr} \ F_{ij} E^i E^j = 0 \eqno(7)$$
Since $A_i$ is complex,
there is also a reality condition that must be
imposed\footnote*{The
 momentum conjugate to $A_i$ is also complex,
but it turns out that its imaginary part commutes with $A_i$. Thus one can
choose $E^i$ to be real. The reality condition is simply that $E^i$ and its
first time derivative - computed via Poisson brackets with the hamiltonian -
be real.}.
The first constraint is the standard  Gauss law constraint of
Yang-Mills theory.
Thus every initial data set for general relativity is
also an initial data set for an SU(2) Yang-Mills theory. The only difference
is that it is also subject to
four additional constraints which are related to reparameterization invariance.
Note that the degrees of freedom match:
SU(2) Yang-Mills theory has $3 \times 2 = 6$ degrees of freedom at each point
which are
reduced to $2$ by the four additional constraints. It should be
emphasized that even though the initial data is  similar, the hamiltonian
for general relativity is very different than Yang-Mills theory. As in the
three dimensional case, reparameterization invariance ensures that
the hamiltonian for general relativity is just a multiple of
the constraints (up to a surface term at infinity).

Notice that all the constraints are simple polynomials in the basic fields.
(The reality condition is also polynomial.)
This is one of the main reasons that this approach initially attracted so much
attention. But just from the form of the constraints it is difficult to
tell how much of an advance this represents. Polynomial equations do not, of
course, imply that the quantum theory
is necessarily
solvable (or even exists!) Although the standard constraints in terms of the
spatial metric and conjugate momentum are not
polynomial, they can be made so by simply multiplying by appropriate
powers of  the determinant of the metric.
Furthermore, the constraint (7) is quadratic
in the momenta, which means that the corresponding operator involves
functional derivatives at the same point and must be regulated. This was
also true in the old variables and was perhaps the main difficulty
in finding solutions to the quantum constraints.
To see the real advantage of this form of the constraints one must
begin to construct the quantum theory.

Classically, the Gauss law
constraint (5) generates gauge transformations, just as
in any gauge theory. One can show that the vector constraint (6) generates
reparameterizations of the three dimensional surface and the scalar constraint
(7) is related to
reparameterizations of time, or motions of the spatial
surface  in the four
dimensional solution.
To construct the quantum theory, we begin by representing states by
functionals of $A_i$. We wish to turn the classical constraints into quantum
operators by
replacing $E^i$ by $ -i \delta /\delta A_i$ and define physical states to be
those
annihilated by the  quantum constraints. This of course requires a choice
of factor ordering. For the constraints linear in the
momenta, the ordering given in (5) and (6) ensures that the quantum
constraints have a similar action as their classical counterparts.
However the quantum scalar constraint (7) has no direct interpretation
since, as we have already mentioned, it
must be regulated. Jacobson and Smolin have shown\refto{js2} that
there exists a regulated form of this
constraint $C_\delta$
and a class of states $\psi_\g$ (parameterized by  a loop $\g$) such that
$$     \lim_{\delta \rightarrow 0} C_\delta\  \psi_\g = 0 \eqno(8)$$
The regulator is a type of point splitting in which the functional derivatives
are evaluated at different points separated by a distance $\delta$.
The states are just the familiar Wilson loops. Given a  non-self-intersecting
smooth closed
curve $\gamma$, set
$$     \psi_\g(A) = Tr P e^{\oint_\g A} \eqno(9)$$
Roughly speaking, the reason this satisfies the constraint is that
each $\delta /\delta A_i$
brings down a term proportional to the tangent vector
to the curve.  Both of these
tangent vectors are contracted into the antisymmetric $F_{ij}$ and hence
vanish.
This type of solution is possible only if one uses variables like Ashtekar's
in which the momentum $E^i$ has two type of indices (the tangent space index
$i$
and an internal index which we have suppressed). Although the constraint (7)
is symmetric under interchange of the two momentum (as it must be), it is
anti-symmetric
 under interchanging each type of index separately.

There are several reasons why one might feel uneasy about this result.
First, since one must introduce a notion of distance to regulate the
constraint,
the regulator breaks three dimensional reparameterization invariance. Formally,
this invariance is restored as the regulator goes to zero, but there is
always the possibility of anomalies. A related difficulty is that $C_\delta
\psi_\g \ne 0$ for $\delta \ne 0 $. Thus in some sense
the regulator breaks four dimensional
reparameterization invariance as well. Finally, the regulated constraint is
not unique. At the moment, there are several proposals for the regulated
constraint which appear to be inequivalent\refto{blencowe}.

Nevertheless, this is a significant achievement. Despite extensive work
on the old canonical
formalism for general relativity, no one has ever achieved an
analogous result. The analog of the scalar constraint in the old variables
is known as the Wheeler-DeWitt
equation. Because of the difficulty in regulating and solving this equation,
extensive work was done on simpler ``minisuperspace" models in which one
freezes out all but a finite number of degrees of freedom of the
metric\footnote*{There is a striking similarity between the motivation
that used to be given
for working on minisuperspace models and the motivation one currently
hears for two dimensional gravity.}.
The full Wheeler-DeWitt equation has never been solved.

Even more remarkable is the fact that the solutions to the (analog of the)
Wheeler-DeWitt equation are just the simple Wilson loops. These have long
been considered
as natural gauge invariant variables for describing Yang-Mills theory both
classically and quantum mechanically\refto{wloops}. The fact that these same
objects solve the scalar constraint of quantum general relativity is
quite surprising.

Although $\psi_\g$ solves the scalar and Gauss law constraint,
it does not solve the
vector constraint. In a key development, Rovelli and Smolin showed that
one can obtain solutions to {\it all}
quantum constraints
by passing to a new representation in which states are functionals of
loops\refto{knots}. This can be obtained formally by the integral
transformation
$$   \psi(\g) = \int {\cal D}A \ W(\g,A) \   \psi(A) \eqno(10)$$
where the kernel is again the Wilson loop
$$     W(\g, A) =  {\rm Tr} \ P e^{\oint_\g A} \eqno(11)$$
This transforms functionals of $A$ into functionals of loops. Alternatively,
the loop representation can be introduced directly by starting with an
algebra of loop observables,
computing their Poisson bracket, and introducing operators on functionals
of loops with the same commutation relations. One then expresses the
constraints as operators in the loop representation. Since gauge invariance
is automatic, there is no analog of the Gauss  constraint.
{}From the above discussion, it might appear that there should be no analog of
the
scalar constraint either. This would indeed be the
case if one could restrict to only smooth non-self-intersecting loops. On the
one hand this sounds reasonable since all gauge invariant information in the
connection is contained in the Wilson loops for this class of $\g$. On the
other hand, to obtain a closed Poisson bracket algebra for the loop
observables,
it seems necessary to work with the larger space of
all piecewise smooth loops\footnote\dag{The Poisson
bracket is nonzero only when two loops intersect, and involves
loops with corners and intersections which result from combining
the original loops.}. Fortunately, even in this larger space, one can
satisfy the scalar constraint by simply restricting the functionals to have
support on just the smooth non-self-intersecting loops.
We can now impose the vector constraint.
This says that the states are invariant under diffeomorphisms of the three
surface.
By definition, the diffeomorphism
class of a smooth non-self-intersecting loop is called a ``knot".
Thus one is led to the remarkable result that {\it functions of knot classes
satisfy
all the constraints of quantum general relativity\refto{knots}!}
This result probably represents the main
achievement of Ashtekar's program so far\footnote{*}{There have
also been applications of these variables
to problems in classical general relativity which we will not review here.}.

In order to be sure that these knot states are physical we must check that
they are normalizable. This is nontrivial since the inner product cannot
be chosen arbitrarily but must be chosen so that physical observables are
hermitian.
(The classical reality condition will enter here.)
Unfortunately, at the present time, very few observables are known
and hence  the inner product has not yet been determined. One might worry
that since one starts with functions on the infinite dimensional space
of loops (or connections) the inner product will necessarily
be a functional integral
which could only be evaluated perturbatively. This would violate the whole
spirit of this nonperturbative approach to quantization. However, one
only needs the inner product on the solutions to the constraints
which, like the knot states above, might well have a countable basis.
In simpler models such as three dimensional gravity\refto{a3d}
and the weak field
limit\refto{linear}, the loop representation  and the inner product
have been constructed with the result that the loop states are normalizable.
This lends support to the idea that they will be normalizable in the
full theory as well.

We have not yet discussed the algebra of the quantum constraints.
If one ignores regularization and formally calculates the commutator
of the quantum
constraints, one finds that there exists a choice of factor ordering such
that the algebra closes\refto{aa}. However
before the discovery of the knot states, there was little reason to trust
this result since it was shown\refto{wf} that regularization has an
important effect
on the operator algebra.
The calculation of the regulated constraint algebra has not yet been completed.
But the existence of solutions to all the constraints shows that there can
be no c-number central extension. Either the constraints will close, or
the nonclosure will be a term which annihilates all the knot states.

How might one physically interpret these knot states? Work on this question is
currently in progress. One possibility is the following.
Consider a  collection of
knots defined
as follows. Take a flat metric on $R^3$ and draw three families of parallel
nonintersecting lines
separated by a distance $a$ as shown in Fig. 1. Now  connect the ends at
infinity to form a knot.
(This can be done in many inequivalent ways.)
This collection of knots are called
{\it weaves}\footnote*{Similar two
dimensional weaves were considered by Witten\refto{ww} in his discussion
of the relation between Chern-Simons theory and integrable models in
statistical
mechanics.}.
A state which is one on these weaves and zero  for all other knots
might be interpreted as representing the flat Euclidean metric on $R^3$.
(One should state this more precisely since a knot is diffeomorphism
invariant while a particular flat metric is not. The correct statement
is that  given a flat metric, one constructs a particular representative
of a knot class to describe it. A diffeomorphism acting on the knot
representative describes the new flat metric obtained by applying the same
diffeomorphism to the original metric.)
Preliminary calculations indicate that the physical
spacing between the lines is
determined by the theory to be the Planck length: If one considers the
operator representing the metric and averages it over  scales much larger
than $a$, then the weave state gives the best approximation to the flat metric
for $a$ equal to the Planck length.
It is tempting to conjecture\refto{rev} that
other background  metrics might correspond to topologically different weaves.
Roughly speaking, given any metric on a three manifold, one might associate
a weave consisting of fibers whose tangent vectors form an orthonormal
basis for the given metric.
It is intriguing to see discrete structure
at the Planck scale emerge from the theory.
In the past, many people have referred to the
``fabric of
spacetime". If these ideas are correct, this phrase may have literal and
not just literary meaning!

The loop representation has also been explored for electromagnetism\refto{em}
and
linearized gravity\refto{linear}. It can be constructed either by transforming
the
connection representation or directly in terms of loop observables.
In this case it suffices to consider simple, unknotted loops.
The result is that for electromagnetism, a one photon state with
momentum $k$ and polarization $\epsilon$ is described by the following
functional of loops:
$$  \psi(\gamma) = \oint d\sigma \epsilon_j \dot \gamma^j e^{ik\cdot \gamma}
	 \eqno(12)$$
This is simply $\psi(\gamma) = \oint_\gamma A$ where $A$ is the  wave function
for the one photon state.
Notice that gauge invariance is automatically enforced by the line integral
around the loop. The states of linear gravity are similar. The main
difference is that for linearized gravity,
there are essentially three copies of the electromagnetic
states (since the self dual connection has three complex internal components).

How does one incorporate these linear states into the full theory?
One possibility
is the following. For each simple loop $\gamma$, one considers a
knot consisting of the  weave together with
the loop $\gamma$ attached (in a manner analogous to ordinary embroidery on
fabric). One now defines a state by the condition that it equal $\psi(\gamma)$
on this knot and similarly for all possible positions  of the loop
in the weave.
Notice that in this picture, gravitons do not make sense on scales less than
the background scale $a$. Any loop smaller than this will be topologically
disconnected.

\subhead{3. OPEN QUESTIONS AND NEW DIRECTIONS}

Although the results that have been obtained so far are promising, there
is much that remains to be done before one can claim to have a consistent
quantum theory of gravity. This section is divided into three parts. In
the first, I consider some open questions in the main program described
above. The second includes a short
discussion of other approaches to quantum gravity using Ashtekar variables.
In the third I consider  the question of whether there exists yet another
set of canonical variables for general relativity (or a theory containing
general relativity) such that the constraints are simplified even further.

\subhead{3.1 Open questions in the main program}

We have already discussed two important unresolved issues in Ashtekar's
approach to quantum gravity. One is to determine
the physical inner product and show that the knot states are
normalizable. The other is to understand better the regularization procedure.
Are there principles which determine it uniquely? Does it lead to anomalies?

One also needs to improve the physical interpretation of the knot states.
For example, can a black hole be described in terms of functionals of knots?
Since there is some indication how to interpret flat space and linearized
gravitons in terms of knot states, one can now begin to consider
graviton-graviton scattering. This will be an important test of this approach
to quantum gravity. Uncontrollable divergences will show that this approach
suffers from the same difficulties of standard quantum field theory methods.
On the other hand, finite answers will be an important confirmation of the
basic principles underlying this approach. A related issue is whether there
exist other solutions to the quantum constraints besides the knot states. If
one can work with a loop representation consisting only of smooth
non-self-intersecting loops, it would appear that the answer is no. If one
works with the larger space of piecewise smooth loops, then additional
solutions
can be found\refto{newsolns}.

So far, I have only considered pure general relativity without matter.
The first step toward including matter is to show that the combined
gravity matter system can be written in terms of the self dual connection
in such a way that the constraints are still polynomial in the basic
canonical variables.
This requires that  the metric and its inverse cannot both appear. This step
has been carried out for scalar, spinor and gauge fields\refto{matter}.
In particular, the action
for supergravity has been written in terms of Ashtekar variables\refto{sugra}.
The next step is to find solutions to the quantum constraints. In the presence
of matter, this is not well understood.
It should be kept in mind that unlike superstring theory, this approach does
not
at present provide
a {\it unified} picture of all forces and matter. Its main advantage (assuming
it is successful) is in staying as close as possible to the experimentally
tested general relativity.

If one considers general relativity with a cosmological constant $\Lambda$,
then one
solution to all of the quantum
constraints turns out to be
$$    \psi(A) = e^{-S_{CS}/\Lambda} \eqno(13)$$
where $S_{CS}$ is the Chern-Simons
action for the self dual connection $A_i$\refto{cs}. The  calculation of the
transform of this state into
the loop representation is similar to the calculation performed by
Witten\refto{wknot}
which reproduced
knot invariants.  One might worry that this indicates that the knot
states will not be normalizable. If one considers the state (13) as a state
in ordinary Yang-Mills theory, then for some choice of $\Lambda$,
it turns out to be
a zero energy eigenstate. But it is outside of the physical Hilbert space
and hence appears to have
no physical significance. Why should the situation for gravity be any better?
The key point is that for gravity one is using self dual connections rather
than real connections. Even for electromagnetism, one can show\refto{ars2}
that if one uses the self dual
representation (and $E^i$ real) then the Chern-Simon's state is just the vacuum
for
one helicity of the photon! (The vacuum for the other helicity is a
constant\footnote*{In the standard treatment one works with positive
frequency fields. Then
self dual configurations describe one helicity and anti-self dual the other.
Here one works only with  self dual fields but allows both positive and
negative
frequency. This explains how both helicities can be obtained and why there is
an asymmetry.}.)
These states are normalizable with respect to the standard Poincare invariant
inner product. This
gives further evidence that the knot states are physical.

As we have mentioned,
Ashtekar's approach to quantum gravity is similar in spirit to the
functional Schroedinger approach to ordinary field theory.
However the reparameterization invariance leads to  the
technical simplification that one does not have to solve the time dependent
Schroedinger equation
since the Hamiltonian is proportional to the constraints.
However this raises a conceptual problem:
How does one recover time and make physical predictions? This is one of the
deep issues that every (nonperturbative) approach to quantum gravity must
address. It
has been discussed extensively\refto{hartle}, but there is still no clear
answer.
Simple models of reparameterization invariant systems suggest that
one part of the argument of the wave function should play the role of time.
In Ashtekar's approach,
there has been some progress in identifying ``time"
in the connection representation\refto{time}
but not much is known yet in the loop representation. Another possibility is
that time will  arise only when gravity is coupled to matter and ``physical
clocks" can be constructed.

\subhead{3.2 Other approaches to quantization with Ashtekar variables}

Although canonical quantization with constraint operators
has been the main focus of work in this
area, it may be worthwhile to examine other approaches. One alternative is
to solve the constraints classically and then quantize the resulting
``true degrees of freedom". (This was in fact the way Witten first
quantized the 2+1 theory.) Remarkably enough, the general solution
of the vector and scalar constraint can be expressed in terms of an
arbitrary symmetric, invertible, traceless $3\times 3$ matrix
$\phi^{ab}(x)$\refto{cdj}.
Given an arbitrary self dual connection $A_i^a$, define
$E^i_a$ so that
$$  F^a_{ij} = \epsilon_{ijk} E^k_b \phi^{ab}  \eqno(14) $$
Substituting this into the constraints, one sees immediately that (6) is
satisfied since $\phi^{ab}$ is symmetric, and (7) is satisfied since
$\phi^{ab}$
is tracefree. One can argue that this is the general solution since $\phi^{ab}$
has five independent components which is the number one expects after solving
four equations for the nine components of $E^i_a$.
Gauss' law is the only remaining constraint on $A_i^a$ and $\phi^{ab}$.
Unfortunately, a simple solution to this equation is not yet available.

Another possibility is to consider covariant approaches to quantization.
This should be more conducive to answering a certain class of questions
such as whether the topology of space can change in quantum gravity.
Even classically, one has the following result.
Both the action  (4) and
the one obtained by replacing $R$ by the curvature of the
self dual connection $A$ do not involve the inverse of the tetrad.
Thus the action and the resulting field equations  remain well defined even
in the limit that the metric becomes degenerate. In general relativity,
it has been shown that any solution to the vacuum Einstein equation which
interpolates between spaces of different topology must be singular. But the
only ``singularity" that is required is for the metric to become degenerate
at one moment of time:
There exist smooth solutions to the equations derived from (4) which change
 topology and have an invertible metric almost
 everywhere\refto{topchg}.

Since Ashtekar's approach and the tetrad approach to general
relativity both naturally
include degenerate metrics, one is faced with the question
of why the metric we see is invertible. In fact, it is not even clear how
to formulate this question precisely. It is tempting to consider the
expectation value of the metric $<g_{\mu\nu}>$, and one often hears
 speculation that $<g_{\mu\nu}>=0$ may correspond to a diffeomorphism invariant
 phase of quantum gravity while $<g_{\mu\nu}>= \eta_{\mu\nu}$ corresponds to
 a state of broken symmetry.
However, it is clear from the quantum constraints that the physical states of
quantum general relativity are
{\it always} diffeomorphism invariant. Moreover,
the expectation value of any non-gauge invariant operator
(such as the metric) must always be gauge invariant: If U denotes a general
gauge transformation, then
$$  <\psi| g_{\mu\nu}|\psi> =  <\psi| U^{-1}  g_{\mu\nu}  U|\psi>
		  \eqno(15)$$
since physical states are gauge invariant.
As there are no nonvanishing diffeomorphism invariant tensor fields, this
shows $<\psi |g_{\mu\nu} | \psi> = 0$ for all physical
states\footnote*{This
assumes that the inner product is defined not just on physical states,  but
also on states such as $g_{\mu\nu} | \psi>$ which are unphysical. Otherwise,
$<\psi |g_{\mu\nu} | \psi>$ is simply not defined.}.
Analogous arguments can be made for spontaneous symmetry breaking in ordinary
gauge theory\refto{elitzer}. However in that case, one can argue that
even though the local symmetry is not spontaneously broken, the corresponding
global symmetry is. It may be possible to extend this argument to gravity
with asymptotically flat boundary conditions. But it certainly cannot
apply to closed universes where
there is no way to disentangle
local and global diffeomorphisms. What is the appropriate gauge invariant
operator which captures the notion of nondegenerate metrics?

\subhead{3.3 Newer variables?}

Although the constraints (5-7) are considerably simpler than the usual form in
terms of the old canonical variables, it is reasonable to ask whether
this is the best one can do. Does there exist an even more clever
choice of variables which will lead to further simplifications? As we
have discussed, one of the constraints in Ashtekar variables is quadratic
in momenta and must be regulated. Are there canonical variables for which
all constraints
are linear in momenta? To see that one's choice of variables can, in principle,
change the
structure of the constraints in this way,
consider again three dimensional general relativity.
In terms of the standard canonical variables (the spatial metric and
extrinsic curvature) the constraints are very similar to the four
dimensional case. In particular, they are quadratic in momenta
(and nonpolynomial in the spatial metric). However
in gauge theory variables, although the constraints are quadratic in the
connection,
they are linear in the triad which is its conjugate momentum.
Thus there is a natural representation in which all constraints are linear
in momentum.

Comparing the actions for general relativity in three (1) and four (4)
dimensions
there are two obvious differences:
the group is
changed from SO(2,1) to SO(3,1) and there is an extra $\e$ in the action.
One can actually
separate these two effects. There are three dimensional
theories
which generalize (1) to any gauge group including SO(3,1). Let $A$ be
the gauge field for an arbitrary Lie group, $F = dA + A \wedge A$
the field strength, and $e$ be a Lie algebra valued one form.
Then one can consider the action\refto{3dtop}
$$    S = \int Tr \ e\wedge F \eqno(16)$$
The constraints are identical to (2) and (3) with $R$ replaced by $F$.
In particular,
they are linear in the momentum conjugate to $A_i$.
In four dimensions, there are theories  with actions similar to (4)
for any gauge group:
$$   S = \int Tr \biggl([e, e] \wedge  F\biggr)  \eqno(17)$$
where $e$ is again a Lie algebra valued one form. For the case of SO(3) the
canonical quantization of this theory has been carried out\refto{hk}. Unlike
the
three dimensional examples, this theory has an infinite number of degrees of
freedom. Nevertheless, once again
all constraints are linear in  momentum.

As a final example,  consider supergravity.
In this theory, the scalar  constraint
can in fact be replaced by its ``square root" - the supersymmetry constraints.
Since the original constraint is quadratic in momentum, one might hope
that the supersymmetry
constraints would be at most linear in the momentum. Unfortunately, this is
not the case.
Although the supersymmetry constraints are linear in the
momentum conjugate to the metric, they contain a term which is the product
of the momentum conjugate to the metric and the momentum conjugate to the
spin 3/2 field\refto{death,sugra}. This is again the product of
functional derivatives at
the same point and must be regulated. In retrospect, it is clear that
the supersymmetry constraints cannot be linear in all momenta: The (anti)
commutator of two constraints linear in momenta is always linear in momenta
and cannot yield the scalar constraint of general relativity.

It is perhaps worth mentioning that going to higher dimensions does
not seem very promising.
In higher dimensions, general relativity can still be expressed in terms
of a Lorentz connection $\w$ and collection of one forms $\e$ with the action
$$  S = \int e^a \wedge \cdots\wedge e^b\wedge R^{cd}\  \epsilon_{a \cdots bcd}
     \eqno(18)$$
However, there is no obvious analog of using the self-dual part of the
connection. Thus it is not even clear how to mimic the simplification
obtained by Ashtekar:
Ashtekar's variables
do not have a natural generalization to higher dimensions.

In conclusion,
I would say that I find the general ideas of Ashtekar's approach to quantum
gravity attractive, and the
results obtained so far intriguing. It is still far from clear whether this
program (or some variation of it)
can be completed, but it certainly seems worth pursuing.

\subhead{ACKNOWLEDGEMENTS}
I wish to thank the organizers of the Strings and Symmetries 1991 Conference
for
a stimulating meeting. I am grateful to A. Ashtekar, C. Rovelli, and
L. Smolin for discussions of their work.  I have  also benefited from
comments and conversations with  S. Giddings, J. Hartle,  T. Jacobson,
M. Srednicki, and A. Strominger.
This work was supported in part by NSF Grant PHY-9008502.

\references

\baselineskip=16pt

\refis{wi3d} E. Witten, \np B311, 46, 1988.

\refis{gs}  M. Goroff and A. Sagnotti, \np B266, 709, 1986.

\refis{rev} For  more complete recent reviews, see
C. Rovelli, ``Ashtekar
formulation of general relativity and loop space non-perturbative quantum
gravity: a report" University of Pittsburgh preprint (1991), to appear in
Class. Quantum Grav.;  A. Ashtekar, {\sl Lectures on Nonperturbative
Canonical Gravity}, (notes prepared in collaboration with R.S. Tate)
World Scientific, 1991.

\refis{fj} See e.g. R. Feynman, \np B188, 479, 1981; R. Jackiw in {\sl
Field Theory and Particle Physics,} World Scientific (1990).

\refis{aa} A. Ashtekar, \prl 57, 2244, 1986; \prd 36, 1587, 1987.

\refis{hu1} J. Charap and M. Duff, \pl B69, 445, 1977; A. Ashtekar and J.
 Samuel, GR12 abstracts, U. of Colorado Press 1989.

\refis{js1} T. Jacobson and L. Smolin, \pl B196, 39, 1987;
\cqg 5, 583, 1988; J. Samuel,
\journal Pramana, 28, L429, 1987.

\refis{sa} J. Samuel, \cqg 5, L123, 1988.

\refis{js2} T. Jacobson and L. Smolin, \np B299, 295, 1988.

\refis{blencowe} See. e.g. M. Blencowe, \np B341, 213, 1990.

\refis{knots} C. Rovelli and L. Smolin, \prl 61, 1155, 1988; \np
B331, 80, 1990.

\refis{sugra} T. Jacobson, \cqg 5, 923, 1988.

\refis{wf} N. Tsamis and R. Woodard, \prd 36, 3641, 1987; J. Friedman and
I. Jack, \prd 37, 3495, 1988.

\refis{cs} H. Kodama, \prd 42, 2548, 1990.

\refis{3dtop} G. Horowitz, \cmp 125, 417, 1989; M. Blau and G. Thompson,
\ap 205, 130, 1991.

\refis{death} P. D'Eath, \prd 29, 2199, 1984.

\refis{hk} V. Husain and K. Kuchar, \prd 42, 4070, 1991.

\refis{wloops} S. Mandelstam, \ap 19, 1, 1962; K. Wilson, \prd 10, 247,
1974; J. Kogut and L. Susskind,
\prd 11, 395, 1975; A. Polyakov, \pl B82, 247, 1979;
\np B164, 171, 1979, Y. Makeenko and
A. Migdal, \pl B88, 135, 1979; G. 'tHooft, \np B153, 141, 1979.

\refis{hartle} See e.g. {\sl Conceptual Problems of Quantum Gravity}, eds. A.
Ashtekar and J. Stachel, Birkhauser, 1991.

\refis{time} A. Ashtekar, in {\sl Conceptual Problems of Quantum Gravity}
(op. cit.).

\refis{ars2} A. Ashtekar,  C. Rovelli, and L. Smolin, ``Self Duality
and Quantization", {\sl J. Geom. Phys.} in press.

\refis{em} A. Ashtekar and C. Rovelli ``A Loop
Representation for the Quantum Maxwell Field", Syracuse preprint (1991).

\refis{linear} A. Ashtekar, C. Rovelli, and L. Smolin, ``Gravitons and Loops",
{\sl Phys. Rev. D.}, to appear.

\refis{ww} E. Witten, \np 322, 629, 1989.

\refis{cdj} R. Capovilla, J. Dell, and T. Jacobson, \prl 63, 2325, 1989.

\refis{topchg} G. Horowitz, \cqg 8, 587, 1991.

\refis{abj} A. Ashtekar, A. Balachandran, and S. Jo, \journal
International Journal of Modern Physics A, 4, 1493, 1989.

\refis{newsolns} V. Hussain, \np B313, 711, 1989; B. Brugmann  and J. Pullin,
``Intersecting N Loop Solutions of the Hamiltonian Constraint of Quantum
Gravity", Syracuse preprint (1990).

\refis{matter} A. Ashtekar, J. Romano, and R. Tate, \prd 40, 2527, 1989.

\refis{a3d} A. Ashtekar, V. Husain. C. Rovelli, J. Samuel, and L. Smolin,
\cqg 6, L185, 1989.

\refis{wknot} E. Witten, \cmp 121, 351, 1989.

\refis{elitzer} S. Elitzur, \prd 12, 3978, 1975.

\endreferences

\subhead{FIGURE CAPTION}

A weave may be interpreted as representing a flat metric
on $R^3$.

\endit
\end